\documentstyle[aps,amssymb]{revtex}

\begin{document}
\author{B. Goodman\thanks{%
goodman@physics.uc.edu} and R.A. Serota\thanks{%
serota@physics.uc.edu}}
\address{Department of Physics\\
University of Cincinnati\\
Cincinnati, OH\ 45221-0011}
\title{Polarizability and Absorption of Small Conducting Particles in a
Time-Varying Electromagnetic Field}
\date{First draft 09/15/99}
\maketitle

\begin{abstract}
We study small conducting particles and thin films in an oscillating
longitudinal electric field. We find the charge, current, and field
distribution in the particle, the polarizability and the electric dipole
absorption. We account for Thomas-Fermi screening by adding a Fick's
diffusion term to Ohm's law. Alternatively, we describe a particle as a
dielectric body with a non-local dielectric constant which is derived in a
microscopic linear-response theory. We show that both approaches are
equivalent.

\begin{enumerate}
\item[PACS:]  73.23.Ps, 78.90.+t
\end{enumerate}
\end{abstract}

\section{Introduction}

The classical theory of polarizability and absorption of small metallic
particles in a time-varying electromagnetic field is a well studied and
documented subject\cite{LL},\cite{BH}. The corresponding quantum theory has
also been studied, starting with the paper of Gor'kov and Eliashberg\cite{GE}
(GE). The quantum treatment can, in principle, address the features of (i)
nonlocality of the current-field relationship, (ii) discreteness and
statistics of the energy levels and (iii) self-consistent screening of the
transition matrix elements. GE assumed that the spectral properties and the
matrix elements were statistically independent and calculated the latter
`semiclassically' for both pure ($\ell >a$) and `dirty' ($\ell <a$) small
particles. They did not take into account the screening of the external
field. This was treated later by Lushnikov, Simonov, and Maksimenko\cite{LMS}
along the same lines as GE.

For dirty systems in particular, the problem has been revisited recently by
Blanter and Mirlin\cite{BM} (BM). They applied the formalism of the
supersymmetric sigma model developed by Efetov\cite{E} for describing the
level statistics and the diffusive nature of the matrix elements of
disordered wavefunctions (see also\cite{E2}). Their work confirmed the
earlier result of GE for unscreened response, thereby justifying GE ansatz
of statistical independence. In a later paper\cite{BM} BM incorporated
screening (at zero frequency only) through an approximate formulation of the
RPA in which the irreducible polarization part was restricted to first order
in the diffusion term.

Mehlig and Wilkinson \cite{MW} have given a phenomenological analysis of the
electric response of small dirty systems. They considered the response to
lowest order in frequency postulating in advance features of the in-phase
and out-of-phase parts of the current relative to the electric field,
namely, that they are ohmic and diffusive, respectively. (We will see later
on that this separation of the current by phase is neither necessary nor
correct in detail.) In a second paper\cite{MW} they restated their earlier
results, starting with the nonlocal conductivity $\sigma \left( {\bf r},{\bf 
r}^{\prime };\omega \right) $ first derived by Serota et. al.\cite{SYK}. As
in the diagrammatic method of Ref.\cite{BM}, the self-consistency equations
for the charge/current/field distribution inside the particle are cast in
integral equation form, which are cumbersome to apply in finite geometries
and have not been solved.

In this paper we show that a considerable simplification in the application
of the semiclassical theory in the above papers to specific geometries is
achieved by modifying the classical Maxwell method for fields in dielectric
and conducting bodies to include nonlocality in the electric
current-electric field constitutive relation. We add to Ohm's law in the
standard Rayleigh-Drude (RD) theory a Fick's law diffusion term, obtaining
what we refer to as the Einstein transport equation. We then follow Landau
and Lifshitz's observation (see\cite{LL}) that an electrically isolated
linear-response system can always be treated as a dielectric body -- in this
case one with a nonlocal dielectric response. However, it will be shown that
even this response function can be avoided. Instead, after formulating
appropriate boundary conditions, we proceed to solve Maxwell's {\it %
differential} equations for the charges and fields in an almost
textbook-like fashion. Thereby self-consistency is achieved at all
frequencies without the cumbersome integral equations of diagrammatic
schemes. The classical treatment of dielectric objects in external fields
is, after all, a self-consistent solution method for charges and their
fields. The great simplicity of this approach relative to the earlier work
mentioned should not distract attention from the fact that it yields without
analytic complication the first exact charge/current distributions for slab
and spherical geometries and that the method is obviously applicable to more
complex geometries.

In Section II we discuss the Maxwell theory with the modified constitutive
equation. The method of solution is presented in Section III, where it is
illustrated for a slab and sphere. Section IV has the discussion of the
equivalence to the earlier nonlocal integral equation first derived by
microscopic theory. In Section V we compare the electric and magnetic dipole
absorption.

\section{The ''dielectric'' equations and boundary conditions.}

In this Section we present the modified RD macroscopic equations for a dirty
metal or semiconductor which includes charge diffusion. The particle is
assumed to be in a uniform quasistatic electric field ${\bf E}=\widehat{{\bf %
z}}E_{0}e^{-i\omega t}$ such as that between parallel condenser plates. This
approximates the electric response to low frequency light where $ka\ll 1$; $%
a $ is the characteristic small dimension of the particle. (The magnetic
response is discussed in Sec. V.) The Maxwell equation for the longitudinal
field and the continuity condition are

\begin{eqnarray}
4\pi \rho \left( {\bf r};\omega \right)  &=&{\bf \nabla \cdot E}\left( {\bf r%
};\omega \right)   \label{Maxwell_eq} \\
i\omega \rho \left( {\bf r};\omega \right)  &=&{\bf \nabla \cdot j}\left( 
{\bf r};\omega \right)   \label{continuity_condition}
\end{eqnarray}
Eliminating $\rho $, we find 
\[
{\bf \nabla \cdot }\left( {\bf j\left( r;\omega \right) -}\frac{i\omega }{4%
\pi }{\bf E}\left( {\bf r};\omega \right) \right) =0
\]
so that the current density can be written as 
\begin{equation}
{\bf j}\left( {\bf r};\omega \right) =-\frac{i\omega }{4\pi }\left( {\bf D}%
\left( {\bf r};\omega \right) -{\bf E}\left( {\bf r};\omega \right) \right) 
\label{j_solution}
\end{equation}
in terms of a divergenceless vector ${\bf D}$ 
\begin{equation}
{\bf \nabla \cdot D}\left( {\bf r};\omega \right) =0  \label{D_eq}
\end{equation}
For simplicity we assume that there is no background dielectric ($\epsilon
_{B}=1$) so that the polarization current is the mobile-charge current ${\bf %
j}$. Then ${\bf D}$ has a meaning of electric displacement field for a
''dielectric'' medium in which all the current is assigned to the
polarization ${\bf P}$, namely, 
\begin{equation}
{\bf j}\left( {\bf r};\omega \right) =-i\omega {\bf P}\left( {\bf r};\omega
\right)   \label{j_vs_P}
\end{equation}
so that 
\begin{equation}
{\bf D}\left( {\bf r};\omega \right) ={\bf E}\left( {\bf r};\omega \right) +4%
\pi {\bf P}\left( {\bf r};\omega \right)   \label{D_full}
\end{equation}
and 
\begin{equation}
{\bf \nabla \cdot P}\left( {\bf r};\omega \right) =-\rho \left( {\bf r}%
;\omega \right)   \label{P_vs_rho}
\end{equation}
The boundary conditions reflect the facts (i) that there is diffusion so
that no infinitely thin surface charge density is present and (ii) that
there is no current through the sample boundary 
\begin{equation}
j_{n}\left( {\bf r};\omega \right) |_{\partial }=0  \label{j_boundary}
\end{equation}
Consequently, 
\begin{equation}
D_{n}\left( {\bf r};\omega \right) |_{\partial }=E_{n}\left( {\bf r};\omega
\right) |_{\partial }  \label{D_boundary_condition}
\end{equation}
In the absence of surface charge, the electric field is continuous at the
boundary 
\begin{equation}
{\bf E}\left( {\bf r};\omega \right) |_{\partial }={\bf E}^{out}\left( {\bf r%
};\omega \right) |_{\partial }  \label{E_boundary_condition}
\end{equation}
where ${\bf E}^{out}$ is the field outside the particle (the sum of the
applied field and the field of the particle). From eqs. (\ref
{E_boundary_condition}) and (\ref{D_boundary_condition}) it then follows
that the normal component of ${\bf D}$ is also continuous at the boundary.
On the other hand, the tangential component of ${\bf D}$ may be
discontinuous due to tangential currents at the boundary.

A\ constitutive equation is needed to complete the description of the
medium. For this we use the\ ''generalized Einstein transport equation'' 
\begin{equation}
{\bf j}\left( {\bf r};\omega \right) =\sigma _{D}{\bf E}\left( {\bf r}
;\omega \right) -D_{D}{\bf \nabla }\rho \left( {\bf r};\omega \right)
\label{Einstein_eq}
\end{equation}
where $\sigma _{D}$ is the Drude conductivity 
\begin{equation}
\sigma _{D}=\frac{\sigma _{0}}{1-i\omega \tau }  \label{sigma_Drude}
\end{equation}
$\sigma _{0}$ and $\tau $ are the Boltzmann conductivity and the scattering
time respectively and 
\begin{equation}
D_{D}=\frac{D}{1-i\omega \tau }  \label{D_Drude}
\end{equation}
where $D$ is the static diffusion coefficient. Thus, the Einstein relation
is taken to hold at $\omega \neq 0$\cite{PN}, namely 
\begin{equation}
\frac{\sigma _{D}}{D_{D}}=\frac{\sigma _{0}}{D}=e^{2}\frac{dn}{d\mu }
\label{Einstein_rel}
\end{equation}
where $dn/d\mu $ is the thermodynamic density of states.

\section{Solution for the field distribution}

\subsection{General Method}

Combining eqs. (\ref{Maxwell_eq}), (\ref{j_solution}), and (\ref{Einstein_eq}
) leads to the following equation for the electric field: 
\begin{equation}
{\bf \nabla }^{2}{\bf E}\left( {\bf r};\omega \right) -\widetilde{\Lambda }%
^{2}{\bf E}\left( {\bf r};\omega \right) =\frac{i\omega }{D_{D}}{\bf D}
\left( {\bf r};\omega \right)  \label{E_eq}
\end{equation}
where $\widetilde{\Lambda }$ is the dynamical screening length given by 
\begin{equation}
\widetilde{\Lambda }^{2}=-\frac{i\omega \epsilon _{D}}{D_{D}}=\Lambda
^{2}\left( 1-\frac{i\omega }{4\pi \sigma _{D}}\right)  \label{Lambda_complex}
\end{equation}
and 
\begin{eqnarray}
\epsilon _{D} &=&1+\frac{4\pi i\sigma _{D}}{\omega }  \label{epsilon_Drude}
\\
\Lambda ^{2} &=&4\pi e^{2}\frac{dn}{d\mu }  \label{Lambda}
\end{eqnarray}
are the Drude dielectric function for a conductor and the static screening
length respectively.

We introduce an auxiliary field ${\bf E}^{h}\left( {\bf r};\omega \right) $
which measures the deviation from the Rayleigh-Drude field relation, namely, 
\begin{equation}
{\bf E}^{h}\left( {\bf r};\omega \right) ={\bf E}\left( {\bf r};\omega
\right) -\frac{{\bf D}\left( {\bf r};\omega \right) }{\epsilon _{D}}
\label{E_solution}
\end{equation}
It obeys the homogeneous PDE 
\begin{equation}
{\bf \nabla }^{2}{\bf E}^{h}\left( {\bf r};\omega \right) -\widetilde{%
\Lambda }^{2}{\bf E}^{h}\left( {\bf r};\omega \right) =0  \label{E_hom_eq}
\end{equation}
and the boundary condition following from eqs. (\ref{D_boundary_condition})
and (\ref{E_boundary_condition}): 
\begin{eqnarray}
E_{n}^{h}|_{\partial } &=&\left( \frac{\epsilon _{D}-1}{\epsilon _{D}}
\right) D_{n}|_{\partial }  \label{E_h_bc} \\
E_{t}^{h}|_{\partial } &=&E_{t}^{out}|_{\partial }  \nonumber
\end{eqnarray}
Equations (\ref{E_hom_eq}) and (\ref{E_h_bc}) determine the field of the
screening charge completely from the boundary values of the displacement
field ${\bf D}$.

Potentials can be introduced since the quasistatic fields are longitudinal: 
\begin{mathletters}
\begin{equation}
{\bf E=}-{\bf \nabla }\phi \text{, }{\bf D=}-{\bf \nabla }\phi ^{d}\text{, }%
{\bf E}^{h}{\bf =}-{\bf \nabla }\phi ^{h}  \label{potentials}
\end{equation}
Then from (\ref{E_solution}) 
\end{mathletters}
\begin{equation}
\phi ^{h}=\phi -\frac{\phi ^{d}}{\epsilon _{D}}  \label{phi_equation}
\end{equation}
Given that the constant in $\phi $ is determined, say, at infinity the
undefined constants in $\phi ^{d}$ and $\phi ^{h}$ must compensate. We will
see below that there is a natural choice for $\phi ^{h}$. The potentials are
determined by solving the Laplace equation for $\phi ^{d}$ in the interior: 
\begin{eqnarray}
\nabla ^{2}\phi ^{d} &=&0  \label{phi_D_equation} \\
\nabla _{n}\phi ^{d}|_{\partial } &=&-E_{n}^{out}|_{\partial }
\label{phi_D_bc}
\end{eqnarray}
plus the auxiliary PDE\ for $\phi ^{h}$: 
\begin{eqnarray}
\nabla ^{2}\phi ^{h}-\widetilde{\Lambda }^{2}\phi ^{h} &=&0
\label{phi_h_equation} \\
\nabla _{n}\phi ^{h}|_{\partial } &=&\left( \frac{\epsilon _{D}-1}{\epsilon
_{D}}\right) \nabla _{n}\phi ^{d}|_{\partial }  \label{phi_h_bc} \\
\left( \phi ^{h}+\frac{\phi ^{d}}{\epsilon _{D}}\right) |_{\partial }
&=&\phi ^{out}|_{\partial }  \nonumber
\end{eqnarray}
$\phi ^{h}$ has the special feature of determining the charge density $\rho $
directly. First, from eqs. (\ref{Maxwell_eq}), (\ref{D_eq}) and (\ref
{phi_equation}) it follows that 
\begin{equation}
{\bf \nabla }^{2}\phi ^{h}=-4\pi \rho   \label{phi_h}
\end{equation}
and hence from (\ref{phi_h_equation}) that 
\begin{equation}
\rho =-\frac{\widetilde{\Lambda }^{2}}{4\pi }\phi ^{h}  \label{rho_vs_phi_h}
\end{equation}
Note that the right hand term of eq. (\ref{phi_h_equation}) is an arbitrary
constant, not necessarily zero. The choice of zero, together with the
vanishing of the total internal charge, fixes the constant in $\phi ^{h}$ to
be such that the average value of $\phi ^{h}$ over the sample is zero.

\subsection{Examples: The field in a sphere and slab}

\subsubsection{Sphere}

This case is presented first because it better represents the method of
potentials just described. Following Ref.\cite{LL} we start with the
required form for the external potential of a dipole 
\begin{equation}
\phi ^{out}\left( {\bf r};\omega \right) =-{\bf E}_{0}{\bf \cdot r}+\frac{%
\overrightarrow{{\cal P}}{\bf \cdot r}}{r^{3}}  \label{phi_out-sphere}
\end{equation}
and $\overrightarrow{{\cal P}}$ ($\parallel {\bf E}_{0}$) is the induced
electric dipole moment of the sphere (of radius $a$). The interior solution
of the Laplace equation (\ref{phi_D_equation}) for $\phi ^{d}$ is
proportional to ${\bf E}_{0}{\bf \cdot r\propto \cos }\theta $ and, from the
boundary condition (\ref{phi_D_bc}), we get 
\begin{equation}
\phi ^{d}\left( {\bf r};\omega \right) =-{\bf E}_{0}{\bf \cdot r}\left( 1+%
\frac{2\alpha }{a^{3}}\right)   \label{phi_in-sphere}
\end{equation}
where $\alpha ={\cal P}/E_{0}$ is the (complex) polarizability. Likewise $%
\phi ^{h}\propto {\bf \cos }\theta $. The solution of (\ref{phi_h_equation})
is 
\begin{eqnarray*}
\phi ^{h}\left( {\bf r};\omega \right)  &\propto &i_{1}\left( r\widetilde{%
\Lambda }\right) {\bf \cos }\theta  \\
i_{1}\left( z\right)  &=&\frac{\cosh z}{z}-\frac{\sinh z}{z^{2}}
\end{eqnarray*}
the latter being the spherical Bessel function of the imaginary argument.
The two boundary conditions (\ref{phi_h_bc}) determine the coefficient of
the solution $\phi ^{h}$ together with the value of $\alpha $, namely, 
\begin{equation}
\phi ^{h}\left( {\bf r};\omega \right) =-{\bf E}_{0}{\bf \cdot r}\frac{%
\epsilon _{D}-1}{\epsilon _{D}+2X}\frac{3ai_{1}\left( r\widetilde{\Lambda }%
\right) }{\sinh \left( a\widetilde{\Lambda }\right) r}
\label{phi_h_solution-sphere}
\end{equation}
and 
\begin{equation}
\alpha =a^{3}\frac{\left( \epsilon _{D}-1\right) X}{\epsilon _{D}+2X}
\label{alpha_solution-sphere}
\end{equation}
with 
\begin{equation}
X=1-3\frac{\coth \left( a\widetilde{\Lambda }\right) }{a\widetilde{\Lambda }}%
+\frac{3}{\left( a\widetilde{\Lambda }\right) ^{2}}  \label{X}
\end{equation}
$X$ increases monotonically to unity as $\left| a\widetilde{\Lambda }\right| 
$ increases. This is very similar to the classical polarizability of a
dielectric sphere with the dielectric constant $\epsilon _{D}$ and reduces
to it when the dynamical screening length is small relative to the radius,
i.e. $a\widetilde{\Lambda }$ is large. The electric dipole absorption is
(see Appendix A for a detailed discussion) is given by the imaginary part of
the polarizability: 
\begin{equation}
Q=\frac{1}{2}\omega E_{0}^{2}%
\mathop{\rm Im}%
\alpha   \label{Q_vs_Pcal}
\end{equation}
The familiar RD\ absorption, namely, 
\begin{equation}
Q_{RD}=\frac{3\omega ^{2}a^{3}}{8\pi \sigma _{0}}E_{0}^{2}
\label{Q_RD-sphere}
\end{equation}
is obtained at low frequencies $\omega \ll \sigma _{0}$. Corrections in $%
\left( a\Lambda \right) ^{-1}$ are easily obtained from eq. (\ref{X}) and in
the lowest order

\begin{equation}
Q=Q_{RD}\left( 1-\frac{11}{2a\Lambda }\right)  \label{Q-sphere}
\end{equation}

\subsubsection{Slab}

Both the slab normal and the field are taken along $z$: $L_{z}\ll
L_{x},L_{y}\thicksim \sqrt{S}$, and $-L_{z}/2<z<L_{z}/2$. The
one-dimensionality permits working directly with the fields. By eqs. (\ref
{D_eq}), (\ref{D_boundary_condition}) and (\ref{E_boundary_condition}) $%
D\left( z\right) =const=E^{out}\equiv E_{0}$ and 
\begin{equation}
\frac{d^{2}}{dz^{2}}E^{h}\left( z;\omega \right) -\widetilde{\Lambda }
^{2}E^{h}\left( z;\omega \right) =0  \label{E_h_eq-slab}
\end{equation}
\begin{equation}
E^{h}(-L_{z}/2)=E^{h}(L_{z}/2)=\left( \frac{\epsilon _{D}-1}{\epsilon _{D}}
\right) E_{0}  \label{bc-slab}
\end{equation}
so 
\begin{equation}
E^{h}\left( z;\omega \right) =E_{0}\frac{\left( \epsilon _{D}-1\right) }{
\epsilon _{D}}\frac{\cosh \left( z\widetilde{\Lambda }\right) }{\cosh \frac{
L_{z}}{2}\widetilde{\Lambda }}  \label{E_h_solution-slab}
\end{equation}
And the internal electric field is 
\begin{equation}
E\left( z;\omega \right) =\frac{E_{0}}{\epsilon _{D}}\left( 1+\left(
\epsilon _{D}-1\right) \frac{\cosh \left( z\widetilde{\Lambda }\right) }{
\cosh \frac{L_{z}}{2}\widetilde{\Lambda }}\right)  \label{E_solution_slab}
\end{equation}
The polarizability and absorption given, respectively, by 
\begin{equation}
\alpha =\frac{L_{x}L_{y}L_{z}}{4\pi }\frac{\left( \epsilon _{D}-1\right) }{
\epsilon _{D}}\left( 1-\frac{\tanh \left( L_{z}\widetilde{\Lambda }/2\right) 
}{L_{z}\widetilde{\Lambda }/2}\right)  \label{alpha_solution-slab}
\end{equation}
and 
\begin{equation}
Q=Q_{RD}\left( 1-\frac{3}{L_{z}\Lambda }\right)  \label{Q-slab}
\end{equation}

In summary, incorporating diffusion into the current field relation modifies
the classical treatment of dielectric bodies by replacing the surface charge
density relation $\sigma =D_{n}\left( \epsilon -1\right) /4\pi \epsilon $ by
an additional degree of freedom for the bulk charge density $\rho \left( 
{\bf r}\right) $. In both cases the Maxwell equations boundary conditions
are all that is needed to assure self-consistency of the charge and field
distribution. It is seen that the screening charge decreases exponentially
with distance $d$ from the surface as $\exp \left( -\widetilde{\Lambda }%
d\right) $ resulting in the correction to the RD absorption proportional to
the fraction of the total volume occupied by the screening charge. This is a
small correction for metals where $\Lambda ^{-1}$ is small but may be
significant in semiconductors.

\section{Relation to response-function formalism}

In this Section we make the connection between the classical dielectric
formalism in Sec. II and the quantum mechanical theories mentioned in the
Introduction. As stated earlier, the dielectric method incorporates
nonlocality in both the conduction and screening consistently.

\subsection{Response formalism}

We start with the nonlocal response function for the electric current vs.
the electric field in the form first given in Ref.\cite{SYK} which was
derived from a field theoretical formalism for diffusive motion, namely, 
\begin{equation}
j_{\alpha }\left( {\bf r};\omega \right) =\int \sigma _{\alpha \beta }\left( 
{\bf r},{\bf r}^{\prime };\omega \right) E_{\beta }\left( {\bf r}^{\prime
};\omega \right) d{\bf r}^{\prime }  \label{j_linear_response}
\end{equation}
where, for both ${\bf r}$ and ${\bf r}^{\prime }$ within the sample, 
\begin{equation}
\sigma _{\alpha \beta }\left( {\bf r},{\bf r}^{\prime };\omega \right)
=\sigma _{D}\left( \delta _{\alpha \beta }\delta ({\bf r}-{\bf r}^{\prime })-%
{\bf \nabla }_{\alpha }{\bf \nabla }_{\beta }^{\prime }d\left( {\bf r},{\bf r%
}^{\prime };\omega \right) \right)  \label{sigma_tensor}
\end{equation}
and $d$ is the diffusion propagator (''diffuson'') satisfying the equation

\begin{equation}
{\bf \nabla }^{2}d\left( {\bf r},{\bf r}^{\prime };\omega \right) =-\delta (%
{\bf r}-{\bf r}^{\prime })-\frac{i\omega }{D_{D}}d\left( {\bf r},{\bf r}
^{\prime };\omega \right)  \label{d_equation}
\end{equation}
and the boundary condition

\begin{equation}
{\bf \nabla }_{n}d\left( {\bf r},{\bf r}^{\prime };\omega \right)
|_{\partial }={\bf \nabla }_{n}^{\prime }d\left( {\bf r},{\bf r}^{\prime
};\omega \right) |_{\partial ^{\prime }}=0  \label{d_boundary_condition}
\end{equation}
Outside the sample $d$ vanishes.

We note here two features of eqs. (\ref{j_linear_response})-(\ref
{d_boundary_condition}): Firstly, the conductivity tensor $\stackrel{%
\leftrightarrow }{\sigma }$ with components $\sigma _{\alpha \beta }$ is
longitudinal, producing longitudinal current ${\bf j}$ when ${\bf E}$ is
longitudinal ($=-{\bf \nabla }\phi $). (Furthermore, ${\bf j}_{n}=0$).
Secondly, the coefficient with $D_{D}$ on the right in eq. (\ref{d_equation}%
) is consistent with the appearance of $\sigma _{D}$ in eq. (\ref
{sigma_tensor}) and both are necessary. With them, eqs. (\ref{sigma_tensor})
and (\ref{d_equation}) are the real-space equivalent of the longitudinal
conductivity, eq. (3.145) of Ref.\cite{PN}, in an unbound homogeneous
medium, namely, 
\begin{eqnarray}
\sigma _{L}\left( q,\omega \right) &=&\frac{ine^{2}}{m}\left[ \omega +\frac{i%
}{\tau }-\frac{s^{2}q^{2}}{\omega }\right] ^{-1}  \nonumber \\
&=&\sigma _{0}\left[ 1-i\omega \tau +\frac{iD_{0}q^{2}}{\omega }\right] ^{-1}
\label{PN_eq}
\end{eqnarray}
where $\sigma _{0}=ne^{2}\tau /m$ and $s$ is the fermi-liquid sound velocity 
\begin{equation}
s^{2}=\frac{n}{m}\frac{d\mu }{dn}=\frac{ne^{2}}{m}\frac{D_{0}}{\sigma _{0}}
\label{sound}
\end{equation}
In $q$-space, eqs. (\ref{sigma_tensor}), (\ref{d_equation}) are 
\begin{eqnarray}
\sigma \left( q,\omega \right) &=&\sigma _{D}\left( 1-q^{2}d\left( q,\omega
\right) \right)  \label{sigma_q} \\
d\left( q,\omega \right) &=&\left( q^{2}-\frac{i\omega }{D_{D}}\right) ^{-1}
\nonumber
\end{eqnarray}
and combine to yield eq. (\ref{PN_eq}).

To connect with Sec. II\ we next show that the constitutive relation (\ref
{j_linear_response}) is equivalent to the Einstein transport eq. (\ref
{Einstein_eq}). We need some variants of eq. (\ref{j_linear_response}).
Inserting ${\bf E=-\nabla }\phi $ in eq. (\ref{j_linear_response}), and
integrating by parts gives

\begin{equation}
{\bf j}\left( {\bf r};\omega \right) =i\omega e^{2}\frac{dn}{d\mu }{\bf %
\nabla }\int d\left( {\bf r},{\bf r}^{\prime };\omega \right) \phi \left( 
{\bf r}^{\prime };\omega \right) d{\bf r}^{\prime }
\label{j_linear_response_2}
\end{equation}
Combining the continuity equation (\ref{continuity_condition}) with eq. (\ref
{d_equation}) gives the nonlocal relation for the charge density 
\begin{equation}
\rho \left( {\bf r};\omega \right) =\int \Pi \left( {\bf r},{\bf r}^{\prime
};\omega \right) \phi \left( {\bf r}^{\prime };\omega \right) d{\bf r}%
^{\prime }  \label{rho_linear_response}
\end{equation}
where 
\begin{eqnarray}
\Pi \left( {\bf r},{\bf r}^{\prime };\omega \right) &=&e^{2}\frac{dn}{d\mu }%
\left( -\delta ({\bf r}-{\bf r}^{\prime })-\frac{i\omega }{D_{D}}d\left( 
{\bf r},{\bf r}^{\prime };\omega \right) \right)  \label{Pi} \\
&=&e^{2}\frac{dn}{d\mu }{\bf \nabla }^{2}d\left( {\bf r},{\bf r}^{\prime
};\omega \right)  \label{Pi-2}
\end{eqnarray}
Equations (\ref{rho_linear_response}) and (\ref{Pi}) give the alternative
form 
\begin{equation}
D_{D}\rho \left( {\bf r};\omega \right) =-\sigma _{D}\phi \left( {\bf r}
;\omega \right) -i\omega e^{2}\frac{dn}{d\mu }\int d\left( {\bf r},{\bf r}%
^{\prime };\omega \right) \phi \left( {\bf r}^{\prime };\omega \right) d{\bf %
r}^{\prime }  \label{alt}
\end{equation}
the gradient of which, combined with (\ref{j_linear_response_2}), yields the
differential (quasilocal)\ constitutive ansatz (\ref{Einstein_eq}) in Sec.
II.

The field equation (\ref{E_eq}) is similarly verified by adding Poisson's
equation to eq. (\ref{alt}), namely, 
\begin{equation}
\left( {\bf \nabla }^{2}-\Lambda ^{2}\right) \phi \left( {\bf r};\omega
\right) =\frac{i\omega \Lambda ^{2}}{D_{D}}\int d\left( {\bf r},{\bf r}%
^{\prime };\omega \right) \phi \left( {\bf r}^{\prime };\omega \right) d{\bf %
r}^{\prime }  \label{alt-2}
\end{equation}
This integro-differential form is typically how self-consistency appears in
the microscopic theories mentioned earlier. But taking the gradient of (\ref
{alt-2}) (and using eqs. (\ref{j_linear_response_2}) and (\ref{j_solution}))
again gives a quasilocal equation which rearranges to (\ref{E_eq}). Thus the
dielectric method of Sec. II\ and eqs. (\ref{j_linear_response}) etc. are
equivalent and the latter can be compared with the previous work.

In Ref.\cite{BM} BM derive the unscreened polarizability of a small particle
which, in the present notation, is given by 
\begin{equation}
\alpha _{0\alpha \beta }=\frac{e^{2}}{2}\frac{dn}{d\mu }\int \int \left(
x_{\alpha }-x_{\alpha }^{\prime }\right) \left( x_{\beta }-x_{\beta
}^{\prime }\right) \left[ \frac{1}{V}-\frac{i\omega }{D_{D}}\overline{d}%
\left( {\bf r},{\bf r}^{\prime };\omega \right) S\left( \frac{\omega }{%
\Delta }\right) \right] d{\bf r}d{\bf r}^{\prime }  \label{alpha_0_BM}
\end{equation}
Here, subscript ''$0$'' means unscreened and $S\left( \frac{\omega }{\Delta }%
\right) $ is the level density correlation function with $\omega
=\varepsilon -\varepsilon ^{\prime }$ and $\Delta $ is the average level
spacing. In the semiclassical limit $\omega \gg $ $\Delta $, $S=1$. (BM\
took $\alpha _{0}$ to be isotropic, so their eq. (18) has $\left( 1/3\right)
\left( {\bf r}-{\bf r}^{\prime }\right) ^{2}$). $\overline{d}$ is $d$ less
the zero-mode term $\propto V^{-1}$ (see eq. (\ref{d_eigenfunctions-1}) so
that $\int \overline{d}d{\bf r}=0$.

The unscreened polarizability here result from replacing $\phi \left( {\bf r}%
\right) $ in eq. (\ref{rho_linear_response}) by $\phi _{0}\left( {\bf r}%
\right) =-{\bf r}\cdot {\bf E}_{0}$, with $x$ measured from the ''center of
mass'', so $\int x_{\alpha }d{\bf r}=0$. The induced electric dipole is 
\begin{eqnarray}
{\cal P}_{0\alpha } &=&\int x_{\alpha }\rho _{0}\left( {\bf r};\omega
\right) d{\bf r=}\left\{ {\bf -}\int \int x_{\alpha }x_{\beta }^{\prime }\Pi
\left( {\bf r},{\bf r}^{\prime };\omega \right) d{\bf r}^{\prime }\right\}
E_{0\beta }  \label{P_0_BM} \\
&=&\alpha _{0\alpha \beta }E_{0\beta }  \nonumber
\end{eqnarray}
which is easily shown to agree with eq. (\ref{alpha_0_BM}) in the
semiclassical limit.

BM extended their work via the RPA\ towards a self-consistent theory; but
their calculation were complicated and limited to lower order in the bubble
diagrams, with the result not being self-consistent.

A second comparison is with the papers by Mehlig and Wilkinson\cite{MW} on
absorption in small particles, particularly the second paper in which they
derive their phenomenological formalism using eqs. (\ref{j_linear_response}%
)-(\ref{d_equation}), but with static values $\sigma _{0}$ and $D_{0}$.
Their principal result is an equation of mixed form (their eq. (9) with sign
change), namely 
\begin{equation}
\left( D{\bf \nabla }^{2}+i\omega \right) \rho =-\sigma {\bf \nabla }
^{2}\phi _{MW}  \label{eq_MW}
\end{equation}
Here $\phi _{MW}$ is an {\em effective} potential, taken to be the sum of
the diffusive part $\phi _{static}$ plus an ohmic part $\phi _{dynamic}$
with separate and pre designed frequency dependencies. Eq. (\ref{eq_MW}) was
intended as the basis for a self-consistent theory.

An equation like eq. (\ref{eq_MW}) derives from the operation of $d^{-1}$,
or ${\bf \nabla }^{2}+i\omega /D$, on eq. (\ref{alt-2}), giving 
\begin{equation}
\left( {\bf \nabla }^{2}-\widetilde{\Lambda }^{2}\right) {\bf \nabla }%
^{2}\phi =0=\left( {\bf \nabla }^{2}-\widetilde{\Lambda }^{2}\right) \rho
\label{alt-3}
\end{equation}
The right hand of (\ref{alt-3}) is the time-dependent Fermi-Thomas- Debye
screening equation. If $\widetilde{\Lambda }^{2}$ is separated in its parts
(eq. (\ref{Lambda_complex})) and $\rho $ is replaced by $-{\bf \nabla }%
^{2}\phi /4\pi $ in the first part, the form (\ref{eq_MW}) results - with
the differences mentioned above.

Boundary conditions on $\phi $ must be added for a complete specification of
the solution of the 4th order PDE on the left or, correspondingly, boundary
conditions on $\rho $ on the right. The form of the latter is already
provided in Sec. III by the boundary conditions (\ref{phi_h_bc}) for $\phi
^{h}$.

The low-frequency absorption cross-section can be obtained by expanding eq. (%
\ref{rho_linear_response}) in powers of $\omega $. The result is that, to $%
O\left( \omega ^{2}\right) $, the absorption is given in terms of static
quantities only and no integral equation is needed. An identical expression,
for a special case of sphere, was given by Maksimenko et. al. in Ref.\cite
{LMS}. Our derivation shows that the region of validity is $\omega <\omega
_{T}$. The details are in Appendix A.

\subsection{The nonlocal dielectric function}

The solution method in Sec. III used the auxiliary potential $\phi ^{h}$ to
bypass the need for the non-local dielectric function. Nonetheless, the
properties of the latter are of some interest since $\epsilon $ gives an
alternative approach to calculating the fields and currents. The relevant
dielectric tensor $\stackrel{\leftrightarrow }{\epsilon }\left( {\bf r},{\bf %
r}^{\prime };\omega \right) $ is derived from the definitions (\ref{j_vs_P}%
), (\ref{D_full}); and the constitutive equation is provided by eq. (\ref
{j_linear_response}), giving 
\begin{equation}
D_{\alpha }\left( {\bf r};\omega \right) =\int \epsilon _{\alpha \beta
}\left( {\bf r},{\bf r}^{\prime };\omega \right) E_{\beta }\left( {\bf r}%
^{\prime };\omega \right) d{\bf r}^{\prime }  \label{D_linear_response}
\end{equation}
with 
\begin{eqnarray}
\epsilon _{\alpha \beta }\left( {\bf r},{\bf r}^{\prime };\omega \right) 
&=&\delta _{\alpha \beta }\delta ({\bf r}-{\bf r}^{\prime })+\frac{4\pi i}{%
\omega }\sigma _{\alpha \beta }\left( {\bf r},{\bf r}^{\prime };\omega
\right)   \label{epsilon_tensor} \\
&=&\epsilon _{D}\left( \omega \right) \delta _{\alpha \beta }\delta ({\bf r}-%
{\bf r}^{\prime })-\frac{4\pi i}{\omega }{\bf \nabla }_{\alpha }{\bf \nabla }%
_{\beta }^{\prime }d\left( {\bf r},{\bf r}^{\prime };\omega \right) 
\label{epsilon_tensor-2}
\end{eqnarray}
The properties of $\stackrel{\leftrightarrow }{\sigma }$ discussed above
eqs. (\ref{j_linear_response})-(\ref{d_equation}) imply that $\stackrel{%
\leftrightarrow }{\epsilon }$ is a {\em longitudinal} tensor (which also
guaranties the boundary condition (\ref{D_boundary_condition})).

The eigenfunction expansion of $\stackrel{\leftrightarrow }{\epsilon }$ is a
natural basis for calculations. It is derived from the corresponding
eigenfunctions of $d$, namely, those of the Laplacian which obey the Neumann
boundary conditions: eq. (\ref{epsilon_tensor}) can be written as 
\begin{eqnarray}
{\bf \nabla }^{2}\Omega _{m}\left( {\bf r}\right) +\lambda _{m}^{2}\Omega
_{m}\left( {\bf r}\right) &=&0  \label{Omega_m_equation} \\
{\bf \nabla }_{n}\Omega _{m}\left( {\bf r}\right) |_{\partial } &=&0
\label{Omega_m_bc}
\end{eqnarray}
These form a complete orthogonal set , with real and positive $\lambda
_{m}^{2}$, except for the $m=0,$ ''zero mode'' 
\begin{equation}
\Omega _{0}=V^{-1/2}\text{, }\lambda _{0}=0  \label{Omega_zero}
\end{equation}
The eigenfunction representation of $d$ is 
\begin{eqnarray}
d\left( {\bf r},{\bf r}^{\prime };\omega \right) &=&\sum_{m=0}^{\infty }%
\frac{\Omega _{m}\left( {\bf r}\right) \Omega _{m}\left( {\bf r}^{\prime
}\right) }{\lambda _{m}^{2}-\frac{i\omega }{D_{D}}}  \label{d_eigenfunctions}
\\
&=&\overline{d}\left( {\bf r},{\bf r}^{\prime };\omega \right) +\frac{%
i\omega }{VD_{D}}  \label{d_eigenfunctions-1}
\end{eqnarray}
where $\overline{d}$ is the sum without $m=0$. It satisfies 
\begin{equation}
{\bf \nabla }^{2}\overline{d}\left( {\bf r},{\bf r}^{\prime };\omega \right)
=-\delta ({\bf r}-{\bf r}^{\prime })+\frac{1}{V}-\frac{i\omega }{D_{D}}%
\overline{d}\left( {\bf r},{\bf r}^{\prime };\omega \right)
\label{d_equation_modified}
\end{equation}
and is the Fourier transform of the time-dependent diffuson of Ref.\cite{GE}.

The orthonormal {\em vector} eigenfunctions are 
\begin{eqnarray}
{\bf v}_{m}\left( {\bf r}\right) &=&\lambda _{m}^{-1}{\bf \nabla }\Omega
_{m}\left( {\bf r}\right) \text{; }m\neq 0  \label{v_m} \\
\int {\bf v}_{m}\left( {\bf r}\right) \cdot {\bf v}_{n}\left( {\bf r}\right)
d{\bf r} &=&{\bf \delta }_{mn}\text{; }\widehat{{\bf n}}{\bf \cdot v}%
_{m}|_{\partial }=0  \label{v_m_bc}
\end{eqnarray}
are complete as discussed below for the expansion of longitudinal tensor
operators.

Let ${\bf A}\left( {\bf r}\right) ={\bf A}_{L}\left( {\bf r}\right) +{\bf A}%
_{T}\left( {\bf r}\right) $, longitudinal and transverse parts. The
eigenfunction sum 
\begin{equation}
\delta _{\alpha \beta }^{L}\left( {\bf r},{\bf r}^{\prime }\right)
=\sum_{m\neq 0}^{\infty }v_{m\alpha }\left( {\bf r}\right) v_{m\beta }\left( 
{\bf r}^{\prime }\right)  \label{delta_fun_longitudinal}
\end{equation}
is the longitudinal $\delta $-function in the sense that 
\[
\int \delta _{\alpha \beta }^{L}\left( {\bf r},{\bf r}^{\prime }\right)
A_{\beta }\left( {\bf r}^{\prime }\right) =A_{L\alpha }\left( {\bf r}%
^{\prime }\right) 
\]
The action of $\delta _{\alpha \beta }^{L}\left( {\bf r},{\bf r}^{\prime
}\right) $ on a longitudinal electric field is equivalent to $\delta
_{\alpha \beta }\delta \left( {\bf r}-{\bf r}^{\prime }\right) $. This leads
to the representation 
\begin{equation}
\epsilon _{\alpha \beta }\left( {\bf r},{\bf r}^{\prime };\omega \right)
=\sum_{m\neq 0}^{\infty }\epsilon _{m}\left( \omega \right) v_{m\alpha
}\left( {\bf r}\right) v_{m\beta }\left( {\bf r}^{\prime }\right)
\label{epsilon_tensor_eigen}
\end{equation}
where 
\begin{equation}
\epsilon _{m}\left( \omega \right) =\frac{D_{D}\lambda _{m}^{2}-i\omega
\epsilon _{D}}{D_{D}\lambda _{m}^{2}-i\omega }=1+\Lambda ^{2}\frac{1}{%
\lambda _{m}^{2}-\frac{i\omega }{D_{D}}}  \label{epsilon_m}
\end{equation}
The inverse operator $\stackrel{\leftrightarrow }{\epsilon }^{-1}$ is
similarly expanded, with coefficients 
\begin{equation}
\left( \epsilon ^{-1}\right) _{m}=\epsilon _{m}^{-1}=1-\Lambda ^{2}\frac{1}{%
\lambda _{m}^{2}+\widetilde{\Lambda }^{2}}  \label{epsilon_m_inverse}
\end{equation}
Eq. (\ref{epsilon_m_inverse}) may be useful for calculating the electrical
field within ellipsoidal samples\cite{LL}. For these cases the interior $%
{\bf D}$-field is constant, $D_{\alpha }{\bf =}a_{\alpha \beta }E_{0\beta }$
(sum on $\beta $), where $a_{\alpha \beta }$ are depolarization factors. For
a slab, $a_{\alpha \beta }=\delta _{\alpha \beta }$ and for a sphere $%
a_{\alpha \beta }=\delta _{\alpha \beta }\left( 1+2{\cal P}%
/E_{0}a^{3}\right) $. Then the interior electric field is 
\begin{equation}
E_{\alpha }\left( {\bf r};\omega \right) =\sum_{m}A_{m}\left( {\bf E}%
_{0};\omega \right) v_{m\alpha }\left( {\bf r}\right)  \label{E_vs_E_0}
\end{equation}
where 
\begin{equation}
A_{m}=\frac{1}{\epsilon _{m}\left( \omega \right) }{\bf J}_{m}\cdot {\bf D=}%
\frac{1}{\epsilon _{m}\left( \omega \right) }{\bf J}_{m}\cdot \stackrel{%
\leftrightarrow }{a}\cdot {\bf E}_{0}  \label{A_m}
\end{equation}
and 
\begin{equation}
{\bf J}_{m}=\int {\bf v}_{m}\left( {\bf r}\right) d{\bf r}  \label{J_m}
\end{equation}
The frequency dependence resides entirely in the coefficients $\epsilon
_{m}\left( \omega \right) $. Substituting ${\bf v}_{m\alpha }$ by $\Omega
_{m}$ gives the potential distribution 
\begin{equation}
\phi \left( {\bf r};\omega \right) =\sum_{m}A_{m}\left( {\bf E}_{0};\omega
\right) \Omega _{m}\left( {\bf r}\right)  \label{phi_ellipsoid}
\end{equation}
Eqs. (\ref{E_vs_E_0}) and (\ref{phi_ellipsoid}) are not simple to evaluate
except for a slab where $\epsilon _{m}$ and ${\bf v}_{m}$ have simple form.
The calculation is given in Appendix B. The solution (\ref{E_solution_slab})
is reproduced. We have not carried out the sum analytically for a sphere and
ellipsoid.

\section{Estimates: Degenerate-electron metal vs. nondegenerate semiconductor
}

\subsection{Screening lengths}

Static screening length is $\Lambda ^{-1}$, where for metals and
semiconductors respectively, 
\begin{eqnarray*}
\Lambda _{met}^{2} &=&\frac{6\pi n_{met}e^{2}}{E_{F}}\text{, Thomas-Fermi} \\
\Lambda _{sc}^{2} &=&\frac{4\pi n_{sc}e^{2}}{k_{B}T\epsilon _{0}}\text{,
Debye (}\epsilon _{0}\text{ - dielectric constant)}
\end{eqnarray*}
and 
\[
\frac{\Lambda _{sc}^{-1}}{\Lambda _{met}^{-1}}=\left( \frac{n_{met}}{n_{sc}}%
\frac{3}{2}\frac{k_{B}T}{E_{F}}\epsilon _{0}\right) ^{1/2}
\]
with $n_{met}\sim 10^{22}-10^{23}cm^{-3}$, $n_{sc}\sim 10^{18}$, $\epsilon
_{0}\sim 10$ screening length in semiconductor is $\sim \left( 10-100\right) 
\times $ screening length in metal. For a metal, 
\[
\Lambda _{met}^{-1}\sim 0.1nm
\]
is much less than mean free path $\ell \sim 10nm$ typical for dirty metal
particles. For a semiconductor, 
\[
\Lambda _{sc}^{-1}\sim \left( 3-10\right) nm
\]

It is evident that the correction $\delta =5.5/\Lambda a$ in eq. (\ref
{Q-sphere}) will be too small to observe in metal particles. For a $50nm$
particle 
\[
\delta _{met}\sim 10^{-2} 
\]
On the other hand, for a semiconductor particle of the same radius 
\[
\delta _{met}\sim .3-1 
\]

\subsection{Magnetic dipole absorption}

The dipolar absorption of light is the sum of electric and magnetic parts 
\[
Q_{E}=\frac{1}{2}\omega E_{0}^{2}%
\mathop{\rm Im}%
\alpha _{E}\text{ and }Q_{H}=\frac{1}{2}\omega E_{0}^{2}%
\mathop{\rm Im}%
\alpha _{H} 
\]
In the electric case the field is screened by the surface layer (neglecting
diffusion) by a factor $\sim \omega /\sigma _{0}$, while, in the magnetic
case the currents penetrate to the skin effect depth $\sqrt{2\pi \sigma
_{0}\omega }/c$, which is greater than the particle size. Taking the
expressions from Ref.\cite{LL} \S \S\ 72-73, we get for a sphere the ratio 
\[
\frac{Q_{H}}{Q_{E}}=\frac{8\pi ^{2}}{45}\frac{a^{2}\sigma _{0}^{2}}{\omega
^{2}}\frac{H_{0}^{2}}{E_{0}^{2}} 
\]
In free space the last factor is one. For the metallic particle ($a\sim 50nm$%
) 
\[
\frac{Q_{H}}{Q_{E}}\sim 40 
\]
but for a semiconductor 
\[
\frac{Q_{H}}{Q_{E}}\sim \left( \frac{\sigma _{sc}}{\sigma _{met}}\right)
^{2}\times \text{above} 
\]
which can be much smaller than one.

\section{Conclusions}

The classical Maxwell dielectric theory based on the ''generalized
Einstein'' constitutive equation (\ref{Einstein_eq}) reproduces the results
of the previous semiclassical treatments of the response of a small
''dirty'' conductor to an oscillating electric field in the limit of the
continuous energy spectrum. When the discreteness of energy levels can no
longer be neglected, the quantum theory becomes important. Both classical
and semiclassical treatments require for their validity that the mean free
path $\ell $ be small in comparison with the particle dimension, in the
latter case because the matrix elements between impurity wave functions are
treated as the excitation of the diffusion propagators. In Sec. IV the
diffusion model, exclusive of level statistics, was shown to be equivalent
to the transport equation (\ref{Einstein_eq}).

The real efficacy of the dielectric method is, of course, for the
quantitative treatment of screening which, in both the nonlocal classical
and quantum formalisms, involves having to solve integral equations in 3D
geometries. The nonlocality is taken into account by the auxiliary field $%
{\bf E}^{h}$ which represents the distributions of charge (see eq. (\ref
{rho_vs_phi_h})). Solving for the fields is not significantly more
complicated than for bodies with local dielectric constants.

An unanswered question in the above considerations is the validity of the
screening description at short distances. There are three lengths of
interest: the mean free path $\ell $, the screening distance $\Lambda ^{-1}$
and the particle size $a$. In metallic particles ($\ell \ll a$) typically $%
\Lambda ^{-1}$ is much less than $\ell $ and the transport equation (\ref
{Einstein_eq}) has questionable physical meaning on the scale of $\Lambda
^{-1}$. Nonetheless, in the limit $\Lambda ^{-1}\rightarrow 0$ where all
charge is on the surface, the classical Rayleigh-Drude theory of metallic
particle is obtained. This suggests that the correction for finite screening
in eqs. (\ref{Q-slab}) and (\ref{Q-sphere})\ is still qualitatively correct.
The same criticism applies to the quantum mechanical descriptions in Refs.%
\cite{GE}-\cite{BM} and \cite{MW} as well as to Pines-Nozieres equation (\ref
{PN_eq}). The theory here may provide a more consistent picture for
semiconductor particles where screening lengths are much larger.

It is interesting that, despite the above mentioned physical limitation, the
mean free path itself does not appear explicitly $-$ only scattering time $%
\tau $ in the Drude form $1-i\omega \tau $. Nor does the Thouless frequency $%
\omega _{T}$ appear explicitly. It does determine the applicability of the
low-frequency absorption in Appendix A.

Finally we conjecture that in the extreme quantum region when the levels are
discrete and $\omega $ and/or $T$ $<\Delta $, use of the ''Golden Rule'' and
impurity-averaged spectral quantities like $S\left( \omega /\Delta \right) $
is fundamentally wrong. We will return to this point in later work.

\appendix 

\section{Low-frequency absorption}

First, we review expressions for absorption of energy from an applied
electric field. The local absorption rate is 
\begin{equation}
q\left( {\bf r};\omega \right) =\frac{1}{2}%
\mathop{\rm Re}%
\left\{ {\bf j}\left( {\bf r};\omega \right) {\bf \cdot E}^{*}\left( {\bf r}%
;\omega \right) \right\}  \label{Q_local}
\end{equation}
The integrated absorption rate 
\begin{equation}
Q\left( \omega \right) =\int q\left( {\bf r};\omega \right) d{\bf r}
\label{Q_total}
\end{equation}
can be rewritten for a longitudinal electric field by applying Gauss'
theorem and the continuity equation, 
\begin{equation}
Q\left( \omega \right) {\bf =}\frac{\omega }{2}%
\mathop{\rm Im}%
\left\{ \int \rho \left( {\bf r};\omega \right) \phi ^{*}\left( {\bf r}%
;\omega \right) d{\bf r}\right\}  \label{Q_total-2}
\end{equation}
The symmetry of this form is used in the derivation of eq. (\ref{rho_1})
below.

A less symmetric form for $Q\left( \omega \right) $ in which $\phi \left( 
{\bf r}\right) $ is replaced by $\phi _{0}\left( {\bf r}\right) $, the
applied field in the absence of the sample, is familiar from thermodynamic
arguments (Ref.\cite{LL}) for quasistatic fields. We remark that replacing $%
\phi $ by $\phi _{0}$ is valid at all frequencies because Coulomb forces are
conservative and do no net work in a cycle. We digress briefly to prove this
(probably well-known) statement, namely that the induced field 
\[
\phi _{ind}=\phi {\bf -}\phi _{0} 
\]
does no work. Because the system is isolated electrically, the total induced
charge $\rho _{ind}$ is zero and $\phi _{ind}\sim r^{-2}$ at large
distances. (An initial charge distribution $\rho _{0}$ does not contribute
to $Q$ and is ignored.) So $\rho _{ind}=\rho $ and 
\begin{equation}
{\bf \nabla }^{2}\phi {\bf =\nabla }^{2}\phi _{ind}=4\pi \rho _{ind}
\label{phi_vs_rho_ind}
\end{equation}
with different conditions at ''infinity.'' To see that 
\begin{equation}
\Delta Q=\int 
\mathop{\rm Im}%
\left\{ \rho _{ind}\phi _{ind}\right\} =0  \label{deltaQ}
\end{equation}
integrate by parts over all space. The surface term goes to zero at large $r$%
, and the new integrand $%
\mathop{\rm Im}%
\left\{ \left| {\bf \nabla }\phi _{ind}\right| ^{2}\right\} $ vanishes.

For the small systems considered here $\phi _{0}\left( {\bf r}\right) =-{\bf %
E}_{0}{\bf \cdot r}$ so that 
\begin{equation}
Q\left( \omega \right) =\frac{\omega }{2}%
\mathop{\rm Im}%
\left\{ \overrightarrow{{\cal P}}{\bf \cdot E}_{0}^{*}\right\}  \label{Q}
\end{equation}
where $\overrightarrow{{\cal P}}$ is the induced moment $\overrightarrow{%
{\cal P}}=\int {\bf r}\rho _{ind}\left( {\bf r};\omega \right) d{\bf r}$,
which is $\int {\bf P}\left( {\bf r};\omega \right) d{\bf r}$ for the
dielectric medium (eq. (\ref{P_vs_rho})). For linear response, $%
\overrightarrow{{\cal P}}{\bf =}\alpha {\bf E}_{0}$, (\ref{Q}) becomes 
\begin{equation}
Q\left( \omega \right) =\frac{\omega }{2}\left| {\bf E}_{0}\right| ^{2}%
\mathop{\rm Im}%
\left\{ \alpha \left( \omega \right) \right\}  \label{Q_vs_alpha}
\end{equation}

To lowest order in $\omega $, $Q\propto \omega ^{2}$. A symmetrical form of
the coefficient of $\omega ^{2}$ can be derived from eq. (\ref{Q_total-2})
by expanding $\rho $ and $\phi $ to first order in $\omega $. Write 
\begin{equation}
\rho =\rho _{st}+\omega \rho _{1}\text{, }\phi =\phi _{st}+\omega \phi _{1}
\label{rho_phi_expansion}
\end{equation}
Now $\phi _{st}=\phi ^{h}$ because $\epsilon _{D}\left( 0\right) =\infty $,
so that 
\begin{equation}
\rho _{st}=-\frac{\Lambda ^{2}}{4\pi }\phi _{st}
\label{rho_static_linear_response}
\end{equation}
The product of the static terms is real so it gives zero in (\ref{Q_total-2}%
), as expected, and leaves 
\begin{equation}
Q=\frac{\omega ^{2}}{2}%
\mathop{\rm Im}%
\left\{ \int \left( \rho _{1}\phi _{st}^{*}-\rho _{st}\phi _{1}^{*}\right) d%
{\bf r}\right\}  \label{Q_omega}
\end{equation}
To $O\left( \omega \right) $ eqs. (\ref{rho_linear_response})-(\ref{Pi-2})
give 
\begin{equation}
\rho _{1}=\int \left[ \Pi _{st}\left( {\bf r},{\bf r}^{\prime }\right) \phi
_{1}\left( {\bf r}^{\prime }\right) +\Pi _{1}\left( {\bf r},{\bf r}^{\prime
}\right) \phi _{st}\left( {\bf r}^{\prime }\right) \right] d{\bf r}^{\prime }
\label{rho_1}
\end{equation}
where 
\begin{equation}
\Pi _{st}\left( {\bf r},{\bf r}^{\prime }\right) =\Pi \left( {\bf r},{\bf r}%
^{\prime };0\right) =-e^{2}\frac{dn}{d\mu }\delta \left( {\bf r}-{\bf r}%
^{\prime }\right)  \label{Pi_static}
\end{equation}
and 
\begin{equation}
\Pi _{1}\left( {\bf r},{\bf r}^{\prime }\right) =-\frac{ie^{2}}{D}\frac{dn}{%
d\mu }d\left( {\bf r},{\bf r}^{\prime };0\right)  \label{Pi_1}
\end{equation}
The term linear in $\phi _{1}$ in the brackets of eq. (\ref{rho_1}) is real
so that the absorption, to order $\omega ^{2}$, is determined by $\omega =0$
quantities 
\begin{equation}
\frac{\omega ^{2}}{2}A\int \int \phi _{st}^{*}\left( {\bf r}\right) d\left( 
{\bf r},{\bf r}^{\prime };0\right) \phi _{st}\left( {\bf r}^{\prime }\right)
d{\bf r}^{\prime }d{\bf r}  \label{Q_LMS_BM}
\end{equation}
where 
\begin{equation}
A=\frac{e^{2}}{D}\frac{dn}{d\mu }=\frac{1}{\sigma _{0}}\left( e^{2}\frac{dn}{%
d\mu }\right) ^{2}=\frac{1}{\sigma _{0}}\frac{\Lambda ^{4}}{\left( 4\pi
\right) ^{2}}  \label{A}
\end{equation}
From the eigenfunction expansion of $d\left( {\bf r},{\bf r}^{\prime
};\omega \right) $, eq. (\ref{d_eigenfunctions-1}), we see that $d\left(
\omega \right) \approx d\left( {\bf 0}\right) $ as long as $\omega <D\lambda
_{1}^{2}\sim \omega _{T}$, the Thouless frequency. The Drude frequency
correction $\omega \tau $ to $\sigma _{0}$ and $D$ is negligible at these
frequencies since $\omega _{T}\tau \approx \left( \ell /L\right) ^{2}$. The
latter must be small in order for the constitutive equation (\ref
{Einstein_eq}) to be meaningful physically.

We have checked (\ref{Q_LMS_BM}) for the spherical case (eq. (\ref
{alpha_solution-sphere})). From eq. (\ref{phi_h_solution-sphere}) 
\begin{eqnarray}
\phi _{st}\left( {\bf r}\right) &=&\left( -{\bf E}_{0}\cdot {\bf r}\right)
f\left( r\right)  \label{phi_stat} \\
f\left( r\right) &=&\frac{3a}{r}%
\mathop{\rm csch}%
\left( a\Lambda \right) i_{1}\left( r\Lambda \right)  \label{f}
\end{eqnarray}
$d\left( {\bf r},{\bf r}^{\prime };0\right) $ is the seen from eqs. (\ref
{d_equation}), (\ref{d_boundary_condition}) be the Coulomb Green function
with Neumann boundary conditions 
\begin{equation}
d\left( {\bf r},{\bf r}^{\prime };0\right) =\frac{1}{D}\left[ \frac{1}{4\pi
\left| {\bf r-r}^{\prime }\right| }+\frac{1}{2V}\left( \frac{r^{2}}{3}%
-a^{2}\right) +\sum_{l=1}^{\infty }\frac{\left( l+1\right) \left( rr^{\prime
}\right) ^{l}}{4\pi la^{2l+1}}P_{l}\left( \cos \theta \right) \right]
\label{d_solution}
\end{equation}
where $P_{l}\left( \cos \theta \right) $ is the Lagrange polynomial and $%
\theta $ is the angle between ${\bf r}$ and ${\bf r}^{\prime }$. Only the $%
l=1$ component 
\begin{equation}
d_{1}\left( {\bf r},{\bf r}^{\prime };0\right) =\frac{1}{D}\left( \frac{1}{%
4\pi }\frac{r_{<}}{r_{>}^{2}}+\frac{rr^{\prime }}{2\pi a^{3}}\right) \cos
\theta  \label{d_1}
\end{equation}
(where $r_{<}$ and $r_{>}$ denote the lesser and greater, respectively, of $%
r $ and $r^{\prime }$) contributes to the integral of eq. (\ref{Q_LMS_BM}).
The result 
\begin{eqnarray}
Q &=&\frac{3\omega ^{2}a^{3}}{8\pi \sigma _{0}}\left| {\bf E}_{0}\right|
^{2}F\left( \Lambda a\right)  \label{Q_appendix} \\
&=&Q_{RD}F\left( \Lambda a\right)  \label{Q_appendix-2}
\end{eqnarray}
where 
\begin{equation}
F\left( z\right) =1-\frac{6}{z}\left( \coth z-\frac{1}{z}\right) +\frac{6}{%
z^{2}}\left( \coth z-\frac{1}{z}\right) ^{2}+\frac{1}{2z}\coth z-\frac{1}{2}%
\left( \coth ^{2}z-1\right)  \label{F}
\end{equation}
agrees with the term $%
\mathop{\rm Im}%
\left\{ \alpha \right\} $ (\ref{alpha_solution-sphere}) which is linear in $%
\omega $.

Finally, we compare (\ref{Q_LMS_BM}) with the expression for absorption
obtained by Maksimenko et. al.. in Ref.\cite{LMS} who extended the
semiclassical method of GE\cite{GE} to take screening into account. The
unscreened dipolar energy $-e{\bf r\cdot E}_{0}$ is replaced by a
''screened'' dipole $-e{\bf R\cdot E}_{0}$. The latter is found within the
RPA\ in which the irreducible polarization part has the same form as
polarization itself, namely, having matrix elements between disordered
states. The solution of the RPA\ integral equation, after an average over
states at the Fermi surface and restricting to spherical geometry, has the 
{\em same form} as eqs. (\ref{phi_stat}), (\ref{f}): 
\begin{eqnarray}
\phi _{st}\left( {\bf r}\right) &=&-{\bf R}\left( {\bf r}\right) {\bf \cdot E%
}_{0}  \label{phi_st_LMS} \\
{\bf R\left( r\right) } &=&{\bf r}f\left( r\right)  \label{R}
\end{eqnarray}
The agreement reflects the correspondence between the self-consistency of
the RPA\ and of the Maxwell dielectric theory. (The resulting formula for $Q$
in Ref.\cite{LMS} contains the level correlation function $S\left( \omega
/\Delta \right) $ and reduces to the present form when $S=1$.)

\section{Dielectric tensor in a slab}

The eigenfunctions and eigenvalues of $d\left( {\bf r},{\bf r}^{\prime
};\omega \right) $ are Corresponding to eqs. (\ref{Omega_m_equation}) and (%
\ref{Omega_m_bc}), 
\begin{equation}
\Omega (z,z^{\prime };\omega )=\sqrt{\frac{2}{L}}\cos \left( \frac{m\pi
\left( z+\frac{L}{2}\right) }{L}\right) ,-\frac{L}{2}\leq z\leq \frac{L}{2}
\label{Omega-slab}
\end{equation}
and 
\begin{equation}
\lambda _{m}^{2}-i\frac{\omega }{D_{D}}=\left( \frac{m\pi }{L}\right) ^{2}-i%
\frac{\omega }{D_{D}}  \label{lambda-slab}
\end{equation}
The vector eigenfunctions are 
\begin{equation}
{\bf v}_{m}\left( z\right) =\widehat{{\bf z}}\sqrt{\frac{2}{L}}\sin \left( 
\frac{m\pi }{L}\left( z+\frac{L}{2}\right) \right)  \label{v_m_slab}
\end{equation}
We treat the case in Sec. III where ${\bf E}_{0}||\widehat{{\bf z}}$ so the
vector notation $\widehat{{\bf z}}$ can be dropped. Then ${\bf D}={\bf E}%
_{0} $ and the prescription in eq. (\ref{E_vs_E_0}) gives

\begin{equation}
\frac{E(z;\omega )}{E_{0}}=\frac{4}{\pi }\sum_{n}^{\infty }\frac{1}{2n+1}%
\left( 1-\Lambda ^{2}\frac{1}{\left( \left( 2n+1\right) \pi /L\right) ^{2}+%
\widetilde{\Lambda }^{2}}\right) ^{-1}\sin \left( \frac{m\pi \left( z+\frac{L%
}{2}\right) }{L}\right)  \nonumber
\end{equation}
The summation is easily done and yields the solution (\ref{E_solution_slab}).

\end{document}